\pdfoutput=1

\documentclass[letterpaper, 10 pt, conference]{ieeeconf}  
\usepackage{amsmath,amsfonts}
\usepackage{algorithmic}
\usepackage{algorithm}
\usepackage{array}
\usepackage[caption=false,font=normalsize,labelfont=sf,textfont=sf]{subfig}
\usepackage{textcomp}
\usepackage{stfloats}
\usepackage{url}
\usepackage{verbatim}
\usepackage{graphicx}
\usepackage{cite}
\usepackage{units}
\usepackage{dsfont}
\usepackage{comment}
\usepackage{booktabs}
\usepackage{marginnote}
\usepackage[dvipsnames]{xcolor}
\usepackage{xfrac}
\usepackage{epstopdf}

\usepackage{svg}

\usepackage{amsthm}
 
\newtheorem{theorem}{Theorem}[section]

\IEEEoverridecommandlockouts                              

\newif\ifmargincomments 
\margincommentstrue

\ifmargincomments

\else

\fi

\title{\LARGE \bf Joint Optimization of Charging Infrastructure Placement and Operational Schedules for a Fleet of Battery Electric Trucks}

\author{Juan Pablo Bertucci$^{1}$, Theo Hofman$^{1}$, Mauro Salazar$^{1}$ 
\thanks{This publication is part of the project GTD-Elektrifikatie, made possible in part by the Ministry of Economic Affairs and Climate Policy of the Netherlands.}
\thanks{$^{1}$Department of Mechanical Engineering, Control System Technology,
        Eindhoven University of Technology, 5600 MB  Eindhoven, The Netherlands
        {\tt\small \{j.p.bertucci, t.hofman, m.r.u.salazar\}@tue.nl}}%
}

\begin{document}

\maketitle
\thispagestyle{empty}
\pagestyle{empty}

\begin{abstract}
This paper examines the challenges and requirements for transitioning logistic distribution networks to electric fleets. 
To maintain their current operations, fleet operators need a clear understanding of the charging infrastructure required and its relationship to existing power grid limitations and fleet schedules.
In this context, this paper presents a modeling framework to optimize the charging infrastructure and charging schedules for a logistic distribution network in a joint fashion.
%
 Specifically, we cast the joint infrastructure design and operational scheduling problem as a mixed-integer linear program that can be solved with off-the-shelf optimization algorithms providing global optimality guarantees.
For a case study in the Netherlands, we assess the impact of different parameters in our optimization problem, specifically, the allowed deviation from existing operations with conventional diesel trucks and the cost factor for daily peak energy usage.
We examine effects on infrastructure design and power requirements, comparing our co-design algorithm with planned infrastructure solutions.
The results indicate that
current charging and electric machine technologies for trucks can perform the itineraries of conventional trucks for our case study, but
to maintain critical time requirements and navigate grid congestion co-design
can have a significant impact in reducing total cost of ownership (average 3.51\% decrease in total costs compared to rule-based design solutions). 

\end{abstract}

\section{Introduction}

Electrification of transportation is mandatory for reducing greenhouse gas emissions, improving air quality, and reducing dependence on fossil fuels. Globally, land transportation accounts for nearly 40\% of CO$_2$ emissions \cite{teter2017future} and within the Netherlands for about 20\% of the total \cite{burgmeijer2018}. Within the transportation sector, road freight accounts for approximately 65\% of the world's freight emissions \cite{itf2021} and 20\% of the transportation greenhouse gas emissions in the Netherlands~\cite{cbsopendata}.

\begin{figure*}[!t]
\begin{centering}
\includegraphics[width=\linewidth,trim={5mm 48mm 5mm 48mm},clip]{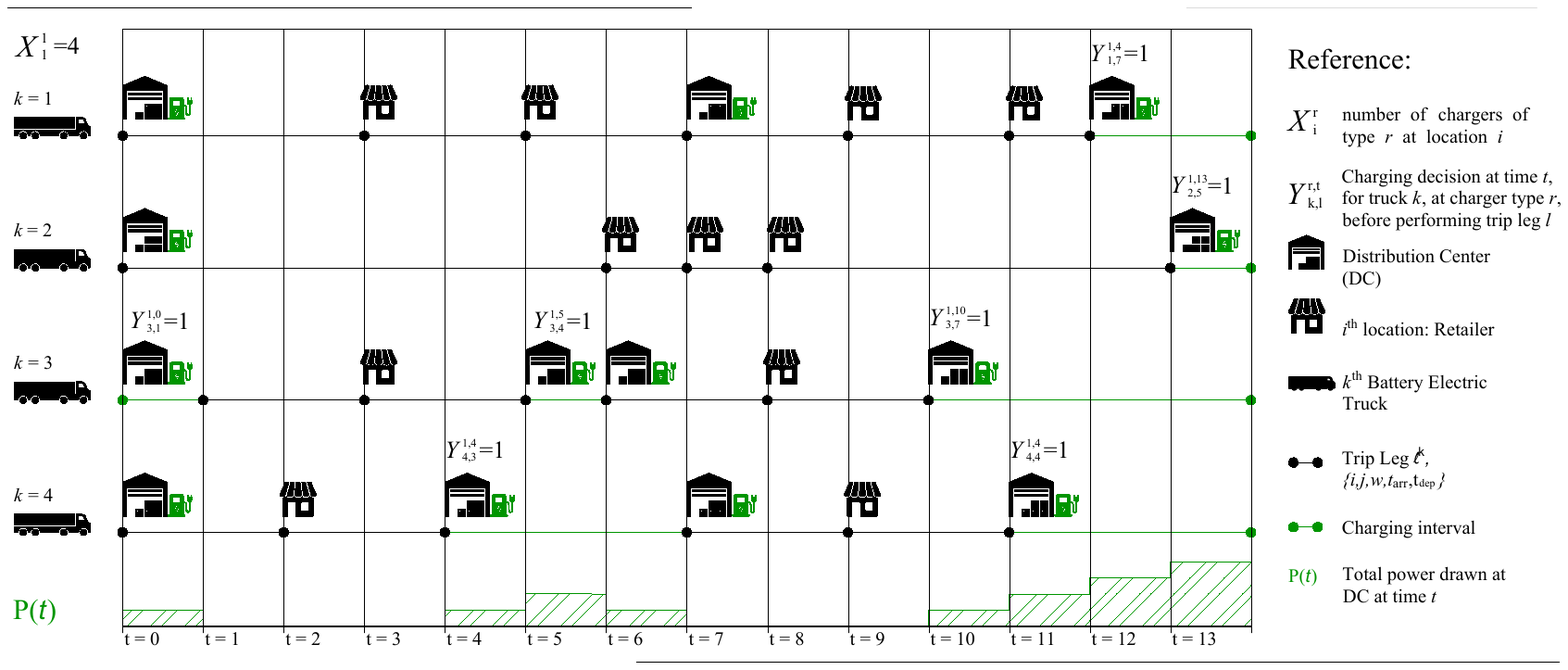}
\caption{
Schematic of the optimization problem solved and its main decision variables. 
$k$ BETs have pre-specified delivery tours comprised of multiple legs $l^k$ that span between locations (DCs or retailers) denoted by $i$ (origin) and $j$ (destination).
The number of chargers of type $r$ available at each location $i$ is given by the integer decision variable $X^{r}_{i}$.
BETs have the decision to charge before each trip leg $l^k$, at a given time step $t$, at a charger of type $r$. This is modelled with a binary decision variable defined as $Y^{k,l}_{r,t}$. 
These decisions add up to a final power requirements curve denoted as $P(t)$. 
}
\label{fig:intro-abm-map}
\end{centering}
\end{figure*}

In response, the Netherlands has put forward an ambitious plan to foment the electrification of freight transport with the placement of Zero Emission Zones (ZEZs) in major cities of the country, where only zero-emission commercial vehicles will be permitted to operate \cite{Rijksoverheid2021Nieuwe}. For instance: The Hague \cite{Den-Haag2021Den}, Rotterdam \cite{Gemeente}, and Amsterdam \cite{Amsterdam0Afspraken} intend to ban the use of commercial non-zero emissions vehicles in and near their downtown areas by as soon as 2025. Additionally, differential tax schemes and purchase investments have been introduced to incentive the purchase of electric trucks in the Netherlands, but also in Germany, Autria, Sweden, the United Kingdom and the United States \cite{Lantz.Joelsson2023}.


This has led to major incentives to transition to battery electric trucks (BETs) for freight, but the challenges of switching to electric trucks are manifold: the trade-off between range (battery size) and useful payload capacity; the lack of charging infrastructure of adequate capacity in critical corridors; and uncertain provision of grid power (especially considering renewable sources) \cite{Ploetz2022,galloElectricTruckBus2016}. All these aspects added to the currently higher purchasing costs compared to traditional diesel vehicles and produce an adverse environment for a quick adoption of electric trucks in supply chains; not to mention possible operational complications due to coordinating the charging of vehicle fleets, whereas the lack of charging infrastructure has been recognized as one of the biggest bottlenecks in the widespread adoption of the technology \cite{Speth.etal2022a,Transport&Environment2023}.

Given the current circumstances, it is imperative to develop new models that guarantee efficient use of assets and energy to facilitate the transition to electric transportation. 
These models will not only accelerate the process but also mitigate potential challenges faced by fleet operators during the transition. The aim of this research is to provide such a model and illustrate the advantages of optimization methods with respect to rule-based infrastructure decision making.



\textit{Literature Review:} Our work contributes to two research
streams, namely, placement of charging stations and scheduling of charging operations for BETs. 
Infrastructure placement for BETs has been tackled with spatial analysis using geographical information systems (GIS), simulation and optimization approaches (as well as combinations of these methods). The location and number of charging points of public fast-charging was obtained in \cite{Speth.etal2022a} using spatial traffic count data and on-site queueing models.
Similarly, \cite{Zhang.etal2021} uses spatial analysis is combined with grid impact models for the case of the US. 
In contrast, spatial analysis of truck trip data is combined with a optimization (solving a facility location problem) in \cite{Whitehead.etal2022}, whereas a %
completely GIS-based approach is given in \cite{Hurtado-Beltran.etal2021}, where existing truck stop facilities along the U.S. Interstate highway system are contrasted with existing vehicle ranges.
Specifically, in \cite{NREL2023} an agent-based model (ABM) was developed to test the performance of individual charging stations. The model provides time-series data on key charging station metrics such as average charging and waiting times, charging loads, among others. In %
\cite{Mishra.etal2022} traffic spatial analysis is combined with the model in \cite{NREL2023} to obtain
the optimal location and power rating of public fast-chargers. In a similar fashion,
\cite{Zhu.etal2021} also employs the model from \cite{NREL2023} and analyzes feasible charging station locations in terms of their impacts on the power grid.
In the case of charging scheduling for BETs, optimization and simulation methods have been used. 
Scheduling is solved with optimization for different combinations of truck and charger characteristics in \cite{pelletierChargeSchedulingElectric2018}, 
and \cite{zhao-vehicle-dispatching} used a gradient-free optimization algorithm to solve an electric vehicle routing problem with pickup and delivery, time windows, and partial recharge for BETs. 
Similarly, \cite{time-capacity} employ genetic algorithms and simulations to determine the optimal charging stop strategies considering a synthetic charging station network.
Later, \cite{bragin-joint} and \cite{wangOptimalDispatchRouting2023a} jointly optimize the routing, fleet size and charging schedule by solving a mixed-integer linear program (MILP) assuming existing charging infrastructure locations.

However, in all these cases the charging infrastructure is assumed or not co-designed with the fleet charging schedules. Moreover, combining this with actual data for the logistics and market-available chargers has not been carried out.
In conclusion, to the best of the authors’ knowledge, no optimization model exists that jointly optimizes charging and route schedules with charging infrastructure for BETs on real-world operations.

\textit{Statement of Contributions:}   In this research, we propose an optimization framework that co-designs the charging infrastructure and charging operations for the existing schedules of traditional diesel-powered trucks. The framework considers current charger and truck technologies being rolled out and the current logistic needs of retailers. This contributes to the emerging literature on BET operations by offering a practical methodology for operators to transition from diesel to electric trucks while efficiently designing their charging infrastructure and operations.

\textit{Organization:} The remaining of this paper is structured as follows: In Section \ref{sec:meth}, we present the optimization program and solution methods used. Section \ref{sec:case} outlines the main variables and experiments deployed in the case study for a distribution center in North Holland. Following that, in Section \ref{sec:res}, we present the results for the case study. Lastly, in Section \ref{sec:dis}, we discuss the methodology, results and directions for future research.


\section{Methodology}\label{sec:meth}

The proposed methodology takes an initial itinerary of an existing logistic operator and solves the problem of determining the optimal number of chargers at each distribution center (DC) and the charging schedule (CS) of each vehicle with a MILP. In this section we describe the required data and problem formulation.

\subsection{Data Inputs}

The main data inputs for the problem are: firstly, the truck fleet data which characterize the BETs that will utilize the charging infrastructure. These details involve the quantity of trucks, their weight, battery size, average speed, and energy consumption. The second data input is the itineraries data set that comprises the routes to be undertaken by the trucks, including their points of origin, destination, distances to be covered, payloads, as well as departure and arrival times. The third type of data, referred to as charger data, pertains to the charging stations that will be used. This includes details of possible locations, rated powers, and efficiencies associated with each charger type. Last, the location data encompasses information about the locations of distribution centers, depots, and retailers where the trucks will be stationed, loaded, unloaded, and charged.

\subsection{Mixed Integer Optimization Problem Formulation}\label{section:opt}
%
We develop a formulation to determine the required infrastructure and optimal charging schedules of a BET fleet for a given set of deliveries to be completed. In this section, we outline the problem formulation as a MILP.

We are provided with an itinerary for each day $d \in \mathcal{D}$ for each truck $k \in \mathcal{K}$ consisting of pickup and deliveries that must be completed  between locations denoted by $i \in \mathcal{I}$ and $j \in \mathcal{J}$.
 We denote each task of a truck $k$ in a day $d$ as a trip leg $l^d_k \in \mathcal{L}^d_k$. We discretize each simulation day $d$ into $t$ units of a size $\tau$ (fraction of hour), such that for each day we have $\frac{d}{\tau}$ time blocks $t$ that span the daily times set $T$ (i.e., $t \in \mathcal{T}$).
 From here onwards we drop the index for day $d$, as the formulation is identical across days, and use simply $l$ to index each element (trip leg) in a tour $\mathcal{L}_k$; thus $l_k$ represents the $l$-th element in the set $\mathcal{L}_k$ (where $l = 1, 2, ..., |\mathcal{L}_k|$)
 For each trip leg there is a required departure time $t^{\mathrm{dep}}_{k,l}$ and arrival time $t^{\mathrm{arr}}_{k,l}$, where the time to complete the trip is given by $t^\mathrm{travel}_{k,l-1}$, and the payload transported between them by $w^k_l$. 
 With these, the tuple $\{i_{k,l},j_{k,l}, t^{\mathrm{arr}}_{k,l}, t^{\mathrm{dep}}_{k,l}, w^k_l\}$ determines the characteristics of an itinerary leg $l$ for a truck $k$. On the infrastructure side, we define the types of chargers as $r \in \mathcal{R}$, which will have different power ratings $e^{\mathrm{char}}_r$ and costs of installation $C^r$.



\subsubsection{Decision Variables}

We allow the vehicles to charge in the time interval between the departure $t^{\mathrm{dep}}_{k,l}$ and arrival $t^{\mathrm{arr}}_{k,l}$ times of a given trip leg $l$. Thus, the choice of a truck $k$ to charge before completing itinerary leg $l$ at a charger of type $r$ at a time slot $t$ can be defined as a binary variable: $Y^{r,t}_{k,l}$. This decision can be taken prior to each departure, to fulfill the energy requirements of performing its upcoming trip. Whether trucks can effectively charge at a location depends on the existing infrastructure, namely, the number of chargers of the required type $r$ at the departure location $i$, denoted as $X^r_i$. The actual time of departure is given by the decision variable $t^{\mathrm{dep.act}}_{k,l}$, which has to be consistent with the chosen charging intervals. 

\subsubsection{Constraints}

The vehicles are subject to energy and available infrastructure constraints:

\textit{Energy Requirements:} We require the vehicles to have enough energy at each origin to arrive at their next destination. For this, we specify the state of energy (SOE) at each destination $j$ of each trip leg $l$ to be at least the energy at the origin $i$ minus the energy need to travel distance $d_{l}$, which we can define by multiplying distance and average consumption per kilometer-ton of each truck $e_k$ as

\begin{equation}
\label{eq:energy-req-veh}
    E^{\mathrm{cons}}_{k,l}  = d^k_l \cdot w_l^k \cdot e_k .
\end{equation}

To be able to make a trip $l^k$, if there is a charger available at the site, a truck may charge at a power rating $e^{\mathrm{char}}_r$. We define the energy charged at the beginning of a leg $l^k$ as

\begin{equation}
\label{eq:energy-charged-pre-trip}
    E^{\mathrm{char}}_{k,l}  = \sum_{t \in T_\mathrm{sub}^{l,k}, r \in \mathcal{R}} Y^{r,t}_{k,l} \cdot e^{\mathrm{char}}_r,
\end{equation}

where the subset $T_\mathrm{sub}^{k,l}$ is defined by the time blocks $t$ that lie between $t^{\mathrm{arr}}_{k,l}$ and $t^{\mathrm{dep}}_{k,l} + \beta^{\mathrm{slack}}$ for a specific truck and itinerary leg. Given that the original itineraries are derived from the vehicle routing of internal combustion engine vehicles, we add the extra parameter $\beta^{\mathrm{slack}}$ which can be tuned to give some additional time slack to the original schedules, such that we can choose to depart from the original schedules if needed for a specific trip. With this, we can then write the first inequality that guarantees the energy feasibility of a trip leg $l^k$ as


\begin{equation}
\label{eq:ctr:energy-leg-update}
    E^{\mathrm{arr}}_{k,l}  \geq E^{\mathrm{dep}}_{k,l}
    + E^{\mathrm{char}}_{k,l}
    - E^{\mathrm{cons}}_{k,l}
\quad \forall l \in \mathcal{L}_k,
\forall k \in \mathcal{K}.
\end{equation}

Additionally, we limit the energy charged to be supported by the vehicle battery $E^\mathrm{max}_k$ at each trip leg:

\begin{equation}
\label{eq:ctr:battery-cap}
E^{\mathrm{dep}}_{k,l} + E^{\mathrm{char}}_{k,l} \leq E^\mathrm{max}_k
\quad \forall l \in \mathcal{L}_k,
\forall k \in \mathcal{K}.
\end{equation}

\textit{Vehicle Schedules:} To have charging schedules that are coherent with the final logistic schedules, we need to enforce that the time of departure is posterior to the latest charging block of trip leg $l$. This is enforced as

\begin{equation}
    \label{eq:ctr:time-coherence-dep-char}
    t^{\mathrm{dep.act}}_{k,l} \geq t \cdot \sum_{r\in R} Y^{r,t}_{k,l} 
    \quad \forall t \in \mathcal{T}_\mathrm{sub},
    \forall k \in \mathcal{K},
    \forall l \in \mathcal{L}_k.
\end{equation}

Similarly, the time of departure chosen and travel time to the next destination has to be consistent with the time window allowed between the actual arrival time from the previous leg $(l-1)$ and the departure time during the current leg $l$:

\begin{equation}
    \begin{split}
    \label{eq:ctr:time-coherence-dep-arr-next}
    t^{\mathrm{dep}}_{k,l} + \beta^{\mathrm{slack}} \geq t^{\mathrm{dep.act}}_{k,l} \geq t^{\mathrm{dep.act}}_{k,l-1} + t^\mathrm{travel}_{k,l-1}
\\ \quad \forall t \in \mathcal{T}_\mathrm{sub},
    \forall k \in \mathcal{K},
    \forall l \in \mathcal{L}_k.
    \end{split}
\end{equation}

\textit{Number of vehicles charging and available chargers:} The number of vehicles charging simultaneously is limited by the total number of available chargers at the departure location.  Thus we write the constraint where for each location $i$ and for each type of charger $r$, we force the total number of vehicles charging at a time interval $t$ to be less than the number of available chargers $X_i^r$:  

\begin{equation}
    \label{eq:ctr:char-ctr}
    X_i^r \geq \sum_{k\in K, l\in \mathcal{L}_k} Y^{r,t}_{k,l} 
    \quad \forall t \in \mathcal{T},
    \forall r \in \mathcal{R},
    \forall i \in \mathcal{I}.
\end{equation}

We also ensure that a vehicle does not charge simultaneously at multiple chargers through

\begin{equation}
    \label{eq:ctr:vehs-unique-char}
    \sum_{r \in \mathcal{R}} Y^{r,t}_{k,l}  \leq 1 
    \quad \forall t \in \mathcal{T},
    \forall k \in \mathcal{K},
    \forall l \in \mathcal{L}_k.
\end{equation}

\subsubsection{Objective}

We define our cost function as the combination of energy, infrastructure, and peak consumption costs:

\textit{Energy Costs:} The time spent charging for each vehicle is given by the time step $\tau$ constant multiplied times the number of time blocks $t$ used for charging. The charging costs are obtained in a similar fashion to the energy charged given at~\eqref{eq:energy-charged-pre-trip}; they are obtained as the amount of power delivered by the charger (given by the power of charge in~\eqref{eq:energy-charged-pre-trip}) multiplied by the energy costs for a charger of type $r$ during time interval $t$: $p^t_r$. We obtain this value for each truck in the fleet:

\begin{equation}
    \label{eq:cost:charging}
     C^{\mathrm{char}}_{k, i} =  \tau \cdot \sum_{t\in T_\mathrm{sub}^{l,k}, r \in \mathcal{R}} Y^{r,t}_{k,l} \cdot \frac{e^{\mathrm{char}}_r}{\eta^r} \cdot p^t_r.
\end{equation}

\textit{Infrastructure Costs:} The infrastructure costs are given by the number of chargers built at each location $X^{r}_{i}$ multiplied by the capital cost of each charger of type $C^r$:

\begin{equation}
    \label{eq:cost:infra}
     C^{\mathrm{infra}}_{i} =  \sum_{r \in \mathcal{R}} C^r \cdot X^{r}_{i}.
\end{equation}

\textit{Peak Consumption Costs:} We define an added cost for peak usage of energy during a day $d$ as the maximum total power charged at any time interval $t$ at a certain location $i$. Thus, we take the maximum value of energy consumption over all time blocks $t$ of a given day to determine $C^{\mathrm{peak}}_{i}$:

\begin{equation}
    \label{eq:cost:peak}
     C^{\mathrm{peak}}_{i} = \max_{t\in\mathcal{T}} \{ \sum_{k \in K, l \in \mathcal{L}_k, r \in \mathcal{R}} \tau \cdot Y^{r,t}_{k,l} \cdot e^{\mathrm{char}}_r \cdot c^\mathrm{peak}\}
\end{equation}

where $c^\mathrm{peak}$ corresponds to the prorrated cost for the peak power used during the entire period of analysis. 
To keep the formulation as a linear problem, we relax the equality ~\eqref{eq:cost:peak} with the following inequality:  

\begin{equation}
    \label{eq:ctr:cost:peak}
     C^{\mathrm{peak}}_{i} \geq \sum_{k \in K,  l \in \mathcal{L}_k, r \in \mathcal{R}} Y^{r,t}_{k,l} \cdot e^{\mathrm{char}}_r \cdot c^\mathrm{peak} \quad \forall i \in \mathcal{I}, \forall t \in \mathcal{T}.
\end{equation}


Since we will be minimizing $C^{\mathrm{peak}}_{i}$ in the objective, it will hold with equality at the solution, yielding an equivalent problem to using ~\eqref{eq:cost:peak}.

Finally, we define the logistic operator's optimization problem to be centered around minimizing the total costs of the operation. This is comprised of the global operation costs (charging and peak costs) and facility construction costs (location, number and type of chargers). Combining ~\eqref{eq:cost:charging}, ~\eqref{eq:cost:infra} and ~\eqref{eq:cost:peak} (multiplied by a parameter $\alpha$, a tuning parameter representing a possible increase in peak power costs) and adding over the itinerary days $d \in \mathcal{D}$ of analysis:

\begin{equation}
    \begin{aligned}
J(X_i^r, Y^{r,t}_{k,l}) =\sum_{d \in \mathcal{D}} \{ \sum_{k \in K, i \in I} C^{\mathrm{char}}_{k, i} 
 + \sum_{i\in I}C^{\mathrm{infra}}_{i} \\
+ \alpha \cdot \sum_{i \in I}C^{\mathrm{peak}}_{i} \}.
    \end{aligned}
\end{equation}

\subsubsection{Problem Formulation}

We can thus write the minimization of the operator's cost as:

\textbf{Problem 1} (Joint Infrastructure Design and Charge Scheduling)
\textit{Given a set of vehicles $k$, with itineraries $L^k_d$, the number and rating of chargers and charging schedules minimizing total cost result from}

\begin{equation}
\begin{aligned}
\label{eq:final-prob}
\min_{X_i^r, Y^{r,t}_{k,l}} \quad & J(X_i^r, Y^{r,t}_{k,l}) \\
\textrm{s.t.} \quad & 
~\eqref{eq:ctr:energy-leg-update},~\eqref{eq:ctr:battery-cap},~\eqref{eq:ctr:time-coherence-dep-char},~\eqref{eq:ctr:time-coherence-dep-arr-next}, ~\eqref{eq:ctr:char-ctr}, ~\eqref{eq:ctr:vehs-unique-char}, ~\eqref{eq:ctr:cost:peak} \\
& Y^{r,t}_{k,l} \in \{0,1\}  \quad \forall r \in \mathcal{R},  t \in \mathcal{T}, k \in \mathcal{K}, l \in \mathcal{L}_k  \\ 
& X_i^r \in \mathbb{N} \quad \forall r \in \mathcal{R}, i \in \mathcal{I}.
\end{aligned}
\end{equation}

Problem 1 is a MILP that can be solved with global optimality guarantees by off-the-shelf optimization algorithms.


\section{Case Study}\label{sec:case}

We analyze a real scenario for a main DC in the Northern Netherlands. We consider three different types of trucks that fulfill the given itineraries: euro trailers, city trailers and rigid trucks, which are typically used for the logistic provision of stores. In this case they are all assumed with the same battery size and consumption per km, but with different carrying capacity.
The scenario optimized uses the itinerary of 3 days of deliveries to 314 locations, with one main DC, where almost all departures occur after 3 AM. Each location had an average daily tonnage delivery of \unit[9.58]{tons}, and an average of 2.15 deliveries were made to each site per day.
For the charging infrastructure we have 5 different chargers with charging powers of 60, 180, 360 and 720 and \unit[1180]{kW}, in accordance with project partner specifications. These have initial costs $C^r$ of 20, 50, 90, 150 and 300 thousand Euro and efficiencies $\eta^r$ of 0.98, 0.98, 0.97, 0.97 and 0.97, respectively. Electricity prices $p_r^t$ and prorrated peak costs $c^\mathrm{peak}$ were obtained for the period of analysis.

We run two distinct explorations, first we analyze the effect of different parameter values ($\alpha, \beta$) of the optimization problem in our infrastructure and schedules solution. Second, we test the value of co-design by benchmarking our joint problem solution to a pre-determined infrastructure planned by project stakeholders for which we optimize the operations only. One of the critical results we want to obtain is the value of jointly optimizing the charging infrastructure with the charging itineraries. We carry out a full factorial exploration across these three dimensions (Peak Factor, Time Slack and Type of Infrastructure Design).

To parse and solve Problem 1, we use Yalmip \cite{Loefberg2004} and Gurobi 10.1 on a Laptop with 16 GB of RAM and an Intel i7-9750H processor. Optimization solutions were obtained within a global optimality gap of less than 1\%.
\subsubsection{Discussion}
We interject at this point a preliminary result; the choice of the parameters was restricted such that the pre-defined infrastructure (rule-based) would produce a feasible problem in the solver; an initial attempt of time slack parameter of 5 minutes could not be carried out by the stakeholder defined infrastructure, whilst it could be produced with the co-design algorithm.
The explanation for this is straightforward, the restriction in the itinerary requires some chargers to be installed in locations outside the main DC, a situation which is not currently envisioned in the pre-defined infrastructure, but accounted for in our methodology.


\section{Results}\label{sec:res}

\begin{figure}[!t]
\begin{centering}
\includegraphics[width=\linewidth]{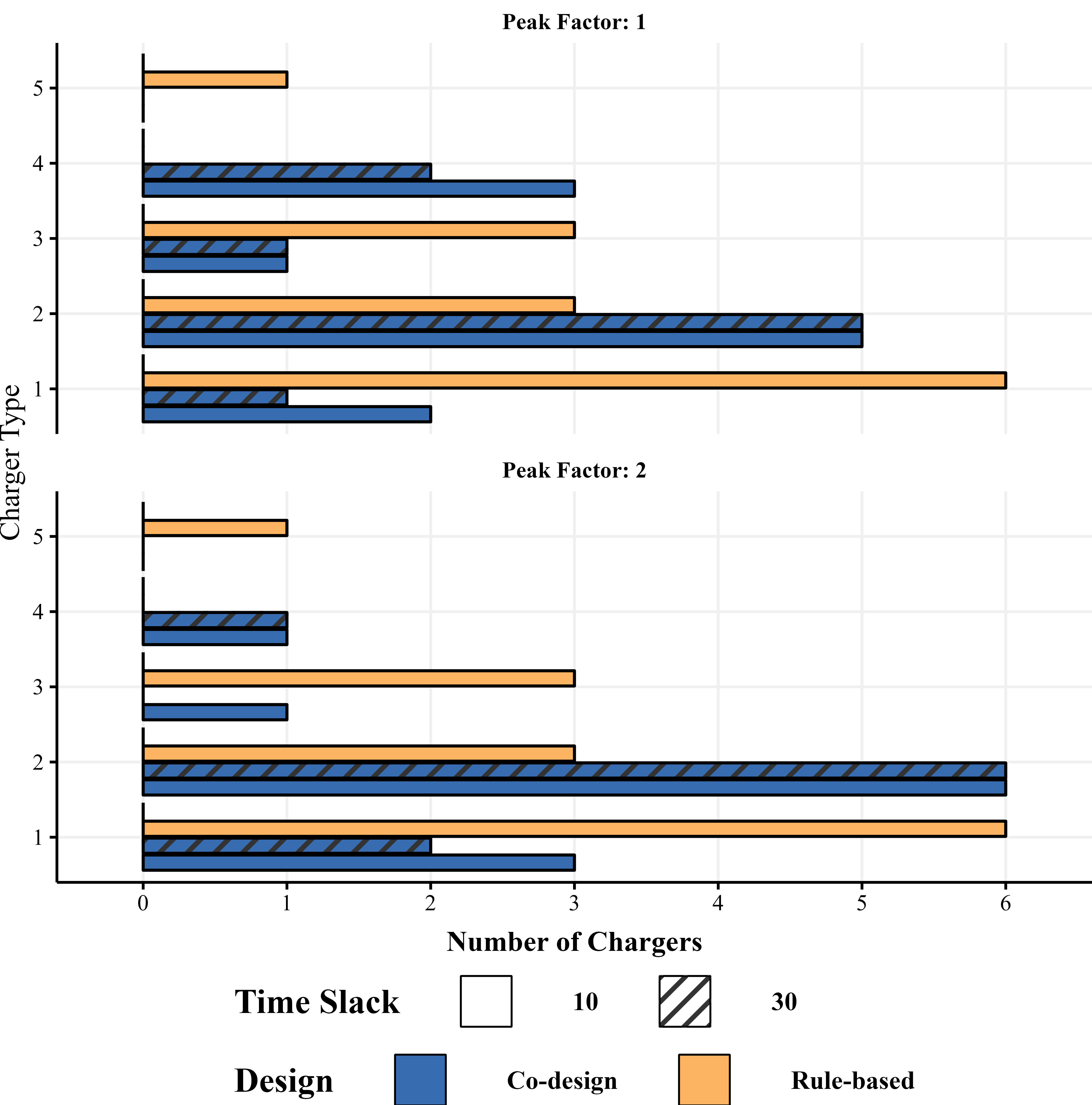}
\caption{Infrastructure results for varying values of time slack and peak factor values. We observe in all cases the use of more Type 2 (\unit[180]{kW}) and Type 4 (\unit[720]{kW}) chargers, and the use of fewer Type 1 (\unit[60]{kW}) chargers. Increasing the Peak Factor induces the use of more, lower-powered chargers in the design.}
\label{fig:res:infra}
\end{centering}
\end{figure}

\begin{figure}[!t]
    \begin{centering}
    \includegraphics[width=\linewidth]{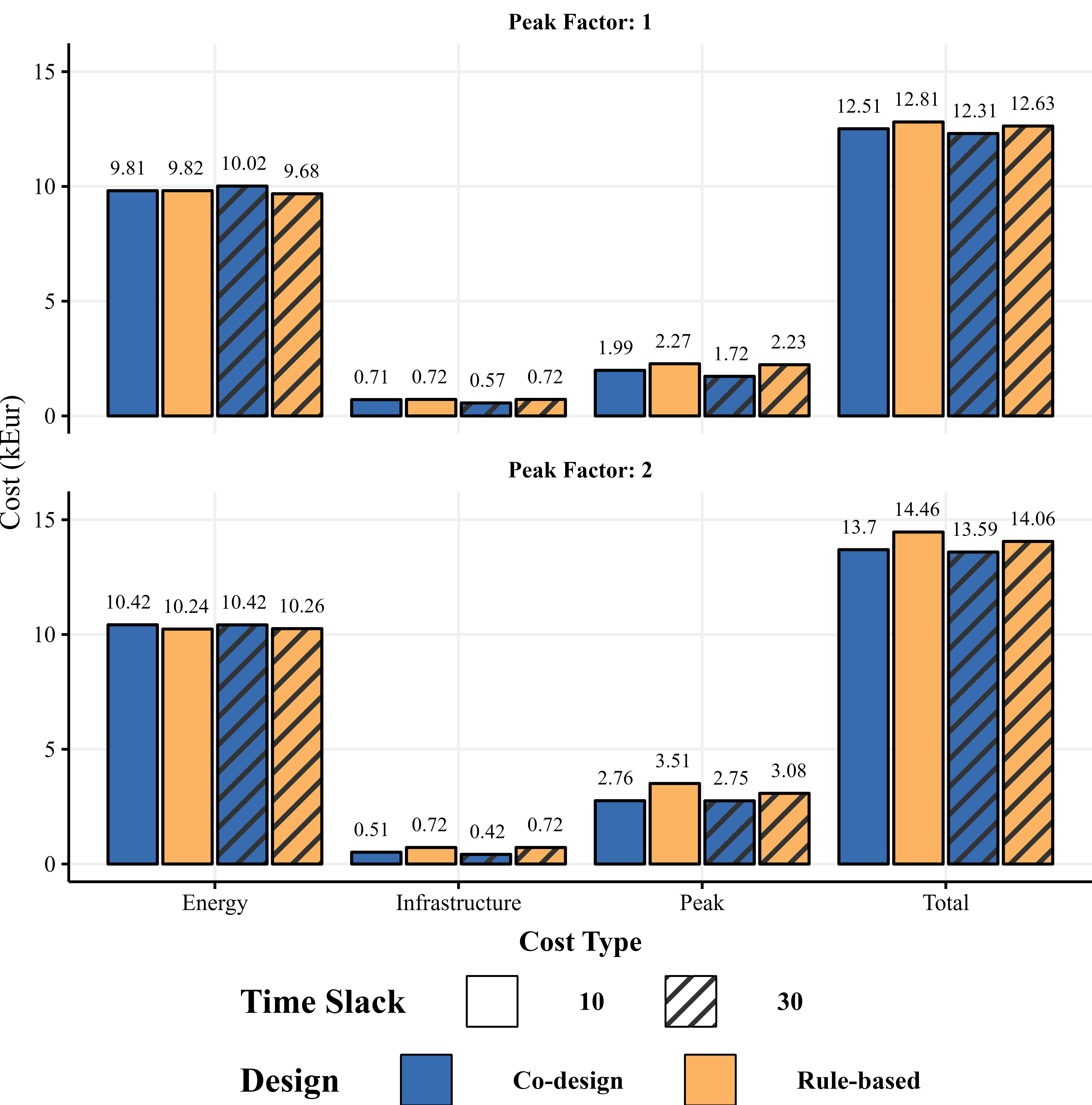}
    \caption{Energy, Infrastructure, Peak and Total costs for each of the experiments. Results are an average of the results for both time slack parameters. Energy costs correspond to the total energy consumed in charging operations weighted by the energy cost at each time $t$. Infrastructure costs are the combined cost of the charger solution amortized 3 days over a 10 year period. The peak costs are proportional to the absolute maximum power for the period of analysis. Total costs are the sum of infrastructure, energy and the peak costs.}
\label{fig:res:costs}
\end{centering}
\end{figure}

\begin{figure}[!t]
\begin{centering}
\includegraphics[width=\linewidth]{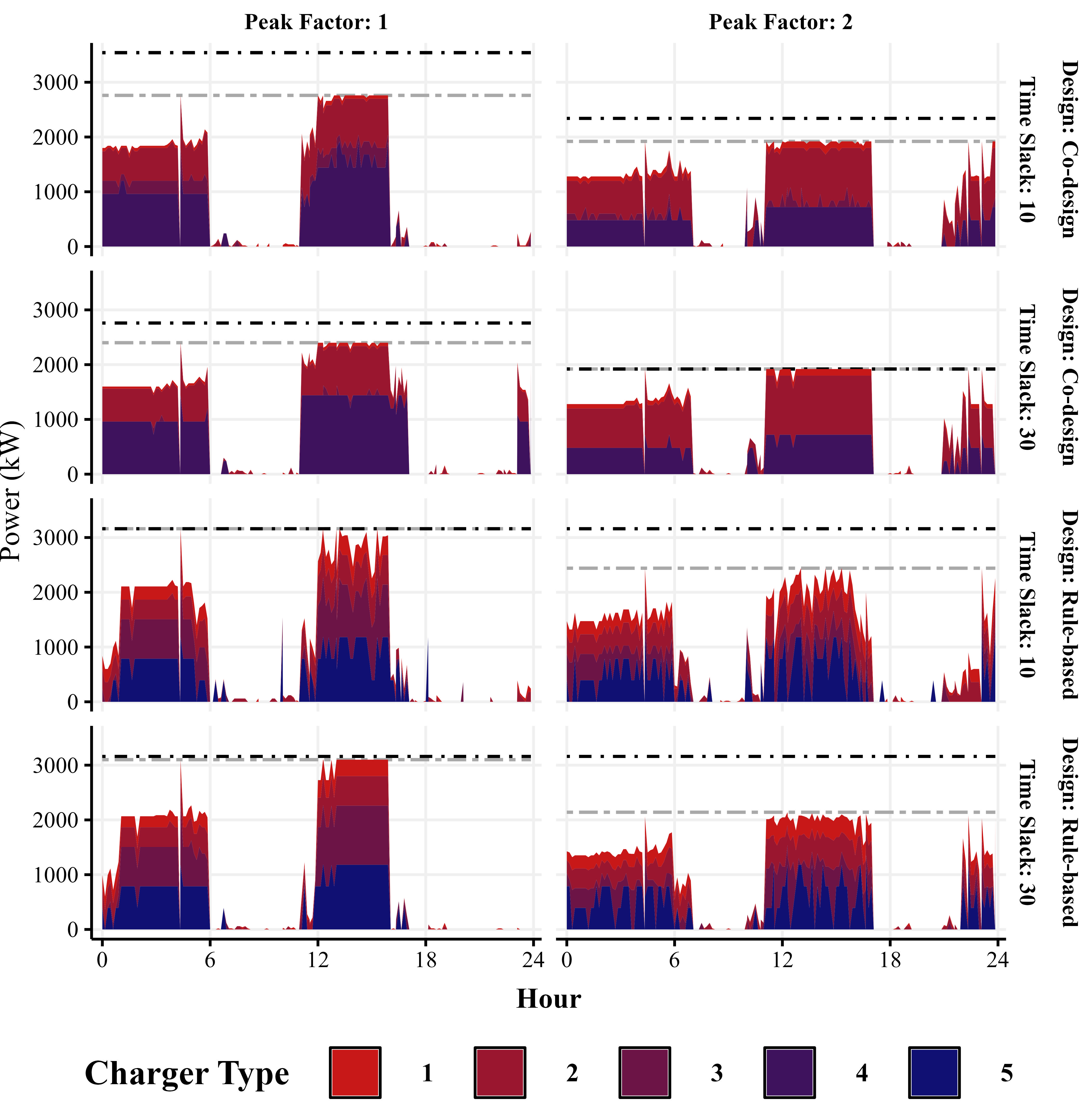}
\caption{Smoothed average daily power consumption curves for the DC by charger type. Results are averaged over the simulation time and given for each peak factor, time slack, and design. The grey dotted line indicates average maximum peak power drawn for the corresponding solution. The black dotted line indicates the maximum installed power of each solution.}
\label{fig:grid-reqs}
\end{centering}
\end{figure}

%
The following section describes the results obtained from the optimization problem for the case study in terms of infrastructure designs in Fig.~\ref{fig:res:infra}, overall costs Fig.~\ref{fig:res:costs}, and in terms of charger usage at the DC in Fig.~\ref{fig:grid-reqs}.

%
As shown in Fig.~\ref{fig:res:infra}, the infrastructure design requires fewer chargers than rule-based infrastructure.
Higher peak factors reduce the used infrastructure, as this means decreased maximum power can be delivered at any time step.
Similarly, increased time slacks allow for more leeway in charging times, and produce a minor reduction in the required chargers. 
In terms of the type of chargers selected, the optimal solutions do not use the highest powered type 5 chargers but rely more on type 4 and 2 chargers. 
The total costs are lower in the co-designed solutions: the reductions are 2.51\% and 4.51\% for the lower and higher peak factor values, respectively (averaging over time slacks).
In general, co-design always yields lower infrastructure costs for this case study (reductions of 14.23\% and 55.72\% ) and savings in peak costs (reductions of 21.88\% and 19.51\%) to the detriment of higher energy costs (increase of 1.67\% and 1.69\%). The availability of more schedule time slacks allows for further savings. 
    %
    %
Thus, co-design for this case study focuses in the reduction of infrastructure and peak power costs in lieu of marginal losses in energy costs; this trade-off follows from the common impact on costs that a reduced infrastructure also has on maximum peak costs, but reduced flexibility to charge during lower energy cost periods.  
    %
%
    %
%
%

These different designs have an impact on the resulting power requirements and charging strategies over the period of analysis.
Following the decreased total infrastructure deployed in co-design, Fig.~\ref{fig:grid-reqs} shows how increased time slacks and peak factors produce lower peak usages.
In all cases the charging is focused on times of lower energy prices, avoiding times between 6-12 and 18-20. The busiest times of operation are between 12 and 18, during which the co-design solutions yield an almost constant type 2 and 4 charger use. 
%
%
%


\section{Conclusion}\label{sec:dis}

In this paper, we presented a framework to jointly optimize the design of the charging infrastructure for BETs, and showcased it in the context of a real-world scenario for the northern Netherlands.
%
%
Our results showed that current schedules with diesel truck fleets can be feasibly performed with BETs and planned infrastructure for our case study when adding at least 10 minutes of extra time to each vehicle itinerary.
The current methodology was able to pinpoint infeasible schedules for a given charging infrastructure, as well as determine the quantity and location of chargers for a given schedule of deliveries. 
It was shown that optimization of charging infrastructure jointly with charging schedules can have a relevant role in reducing both, required infrastructure and operational costs; whilst not co-designing the infrastructure, even if considering an optimized charging schedule, translates to more expensive operations in the long run.
%
%
%
The results also support the development of higher-speed chargers, not only for meeting tight schedules, but also for long-term energy cost savings, whereas for this case the maximum rated power required by co-designed solutions is \unit[720]{kW}.
This work can be extended as follows: First, we plan to validate our strategies via high-fidelity simulations. Second, we would like to generalize our framework to include other means of transportation.

\addtolength{\textheight}{12cm}   


\section*{Acknowledgment}

We thank Dr.~I.~New, F.~Paparella, L.~Pedroso and F.~Vehlhaber for proofreading this paper. This publication is part of the project GTD-Elektrifikatie, made possible in part by the Ministry of Economic Affairs and Climate Policy of the Netherlands.

\bibliographystyle{IEEEtran}

\bibliography{main}

\end{document}